\documentclass[sigconf]{acmart}
\AtBeginDocument{%
  \providecommand\BibTeX{{%
    Bib\TeX}}}

\copyrightyear{2026}
\acmYear{2026}
\setcopyright{cc}
\setcctype{by}
\acmConference[DAC '26]{63rd ACM/IEEE Design Automation Conference}{July 26--29, 2026}{Long Beach, CA, USA}
\acmBooktitle{63rd ACM/IEEE Design Automation Conference (DAC '26), July 26--29, 2026, Long Beach, CA, USA}
\acmDOI{10.1145/3770743.3804145}
\acmISBN{979-8-4007-2254-7/2026/07}

\usepackage{algorithmic}
\usepackage{textcomp}
\def\BibTeX{{\rm B\kern-.05em{\sc i\kern-.025em b}\kern-.08em
    T\kern-.1667em\lower.7ex\hbox{E}\kern-.125emX}}

\usepackage{soul}
\usepackage{xspace}
\usepackage{makecell}
\usepackage{multirow}
\usepackage{graphicx}
\usepackage{subcaption}
\usepackage{listings}
\usepackage{xcolor}
\usepackage{verbatim}
\usepackage{enumitem}

\usepackage{caption}
\usepackage{listings}
\usepackage{bbding}

\usepackage{amsmath}

\usepackage{pifont}

\newcommand{\tool}{{\scshape BluesFL}\xspace}
\newcommand{\algo}{{\scshape Blues}\xspace}
\newcommand{\naive}{{\scshape Agentless}\xspace}

\newtheorem{definition}{Definition}
\usepackage[ruled,vlined,linesnumbered,noend]{algorithm2e}

\begin{document}

\title{Debug Like a Human: Scaling LLM-based Fault Localization to Processor Design via Block-Level Instruction-Oriented Slicing}

\author{Zizhen Liu}
\affiliation{%
  \institution{National University of Defense Technology}
  \city{Changsha}
  \state{Hunan}
  \country{China}
}
\email{zizhen\_liu@nudt.edu.cn}

\author{Xiaoguang Mao$^{*}$}
\affiliation{%
  \institution{National University of Defense Technology}
  \city{Changsha}
  \country{China}
}
\email{xgmao@nudt.edu.cn}

\author{Deheng Yang$^{*}$}
\affiliation{%
  \institution{Academy of Military Sciences}
  \city{Beijing}
  \country{China}
}
\email{dehengyang@outlook.com}

\author{Jiayu He}
\affiliation{%
  \institution{National University of Defense Technology}
  \city{Changsha}
  \country{China}
}
\email{hejy47@nudt.edu.cn}

\author{Yihao Qin}
\affiliation{%
  \institution{National University of Defense Technology}
  \city{Changsha}
  \country{China}
}
\email{yihaoqin@nudt.edu.cn}

\author{Guangda Zhang}
\affiliation{%
  \institution{Academy of Military Sciences}
  \city{Beijing}
  \country{China}
}
\email{zhanggd_nudt@hotmail.com}

\author{Yan Lei}
\affiliation{%
  \institution{Chongqing University}
  \city{Chongqing}
  \country{China}}
\email{yanlei@cqu.edu.cn}

\author{Jianjun Xu}
\affiliation{%
  \institution{National University of Defense Technology}
  \city{Changsha}
  \country{China}
}
\email{jjxu@nudt.edu.cn}

\author{Jiang Wu}
\affiliation{%
  \institution{Academy of Military Sciences}
  \city{Beijing}
  \country{China}
}
\email{wujadeon@outlook.com}

\renewcommand{\shortauthors}{Zizhen Liu, Xiaoguang Mao$^{*}$, Deheng Yang$^{*}$, Jiayu He, Yihao Qin, Guangda Zhang, Yan Lei, Jianjun Xu and Jiang Wu}

\thanks{$^{*}$ Co-corresponding authors: Xiaoguang Mao and Deheng Yang}

\begin{abstract}
Fault localization in modern processor design code is a critical yet time-consuming step during processor verification.
While recent advances in LLM-based techniques for module-level hardware design have shown promising results, automatically localizing bugs in large-scale, project-level processor designs remains challenging.
In this paper, we present \tool, a novel block-level LLM-based fault localization framework for processor designs.
Inspired by the way engineers debug processors, we first propose a dataflow-based code blockization approach to guide LLMs to focus on critical local code context.
We further propose a \underline{B}lock-\underline{l}evel Instr\underline{u}ction-Ori\underline{e}nted \underline{S}licing (\algo) algorithm that enables LLMs to mimic human reasoning by analyzing instruction execution paths and processor states.
We evaluate \tool on a real-world RISC-V processor core comprising 19K lines of SystemVerilog code.
Experimental results demonstrate that \tool correctly localizes 24 bugs at Top-1, achieving 242.9\% improvement over the existing state-of-the-art (7 bugs). Cost analysis shows that \tool requires an average of only \$0.257 to localize a single bug.
\vspace{-5pt}
\end{abstract}




\maketitle

\vspace{-5pt}
\section{Introduction}
Modern processors contain millions of lines of hardware description language (HDL) code to enhance performance and extend functionality. However, this increasing complexity also makes them highly susceptible to design bugs, which can result in severe functional errors \cite{cpushookworld}, degraded performance \cite{barboza2021automatic_cpu_perf}, or even critical security vulnerabilities \cite{lipp2018meltdown, kocher2020spectre}.
Localizing bugs in processor source code is essential for developers to fix them.
However, this debugging process remains one of the most time-consuming stages of processor development.
According to recent surveys, verification-related tasks account for approximately 60\% of the total hardware development effort \cite{Wilson-effort}.
Therefore, automatically locating bugs in HDL source code (e.g., Verilog) for large-scale processor designs is crucial for achieving agile development, as inefficient debugging directly impacts design productivity and undermines market competitiveness.

Various techniques have been proposed to help verification engineers automatically detect processor bugs \cite{formal_verif_dac15, miftah2025symbfuzz, hur2021difuzzrtl, zhou2015functional, wagner2005stresstest}.
However, most methods primarily focus on detecting, leaving the task of locating them within the source code to developers \cite{ma2022debugging}.
Once a bug is detected, engineers typically examine suspicious signals, and manually review the Verilog code to identify the buggy lines.
This manual debugging process is both time-consuming and inefficient.
To alleviate this burden, numerous fault localization (FL) techniques for hardware designs have been proposed.
More recently, large language model (LLM)-based methods \cite{lik} directly predict buggy lines from Verilog modules, demonstrating significant promise.
While existing techniques have proven effective at general or module-level hardware designs,
FL for real-world, project-level processor design code is still at an early stage currently and has the following limitations:

\textbf{Limitation 1: To the best of our knowledge, no specific FL technique for processor functional bugs has been studied yet.}
Processors, as instruction-driven hardware, are inherently more complex and distinctive than general hardware modules (e.g., memory controller, ALU).
Consequently, FL techniques designed for general hardware may be less effective for processors, as assumptions valid for small-scale hardware may not hold in the processor domain.
For example, slicing-based approaches \cite{ahmad2022cirfix, yang2023strider} perform dataflow analysis to produce a set of suspicious lines.
While this set is small enough for most hardware modules to review manually,
it becomes considerably larger in large-scale processor designs, making manual review impractical.
Hence, dedicated techniques are required to achieve effective bug localization in processor designs.


\textbf{Limitation 2: Existing hardware FL techniques lack effective utilization of instruction execution coverage.}
Modern processors use pipelining to execute multiple instructions simultaneously, causing instruction execution to span both spatial and temporal dimensions.
The execution of an instruction from the fetch to the commit stage occurs across multiple modules in space and multiple cycles in time.
Existing spectrum-based hardware FL techniques \cite{spectrum_wu} typically rely on coarse-grained, range-based coverage collection over simulation intervals.
As a result, coverage data for an instruction may be polluted by other in-flight instructions, reducing localization accuracy.
Therefore, accurately capturing the execution path of each instruction is crucial.



\textbf{Limitation 3: Existing hardware FL techniques rarely leverage concrete signal values to aid localization.}
Signal values are essential for human developers to understand hardware behavior~\cite{ma2022debugging}.
However, current hardware FL techniques seldom utilize this information because doing so requires advanced semantic comprehension.
Recent advances in LLMs make it feasible to interpret signal values for reasoning about instruction behavior.
Nevertheless, prompting LLMs with processor waveforms remains challenging.
Processor simulations generate thousands of signals across thousands of cycles, producing extremely large waveforms.
Identifying which signals at which cycles should be provided to LLMs to aid localization remains a nontrivial problem.


\textbf{Limitation 4: LLMs struggle to localize bugs in excessive code context.}
Modern processor projects typically consist of numerous module files whose combined size exceeds the context window limits of LLMs.
When prompted with overly large code contexts, LLMs often lose focus and overlook critical information~\cite{liu-etal-2024-lost}.
Unlike module-level hardware designs, where the entire design can be inspected by the model, processor-scale code introduces substantial noise that hinders the model's ability to identify subtle bugs.
Therefore, providing LLMs with critical and compact code context is essential for enabling accurate fault localization in large-scale processor designs.

To address the above limitations, we derive two key insights from developer debug experience: (1) developers primarily focus on local code regions with dataflow relationships, and (2) they trace instruction execution paths for understanding processor behavior.
Motivated by these observations, we first propose a dataflow-based code blockization method that preserves local semantics while maintaining compact code size (\textbf{Limitation 4}).
Building upon this foundation, we introduce \tool, a LLM-based block-level fault localization framework for large-scale processor designs (\textbf{Limitation 1}).
\tool mimics human debugging behavior by reviewing instruction execution paths and assigning suspiciousness scores to buggy code blocks.
Specifically, a novel Block-Level Instruction-Oriented Slicing (\algo) algorithm is proposed to guide LLMs to focus on instruction-executed blocks (\textbf{Limitation 2}) and enhance their understanding of instruction behavior by incorporating the values of signals through time-annotated blocks (\textbf{Limitation 3}).

In summary, this paper makes the following contributions:
\vspace{-3pt}
\begin{itemize}[left=0pt]
    \item We propose \tool, a block-level LLM-based fault localization framework for processor designs. To the best of our knowledge, this is the first approach that applies LLM-based fault localization to real-world processors.
    \item We introduce a dataflow-based code blockization method that preserves local semantics while producing compact blocks for efficient LLM analysis.
    \item We propose a Block-Level Instruction-Oriented Slicing algorithm that helps LLMs understand instruction dynamic behavior.
    \item We evaluate \tool on a 19KLoC open-source RISC-V core, where it correctly localizes 24 bugs at Top-1. All source code and results are available at: \textcolor{blue}{\url{https://github.com/pointerliu/bluesfl}}.
\end{itemize}
\vspace{-10pt}

\section{Background}

\subsection{Co-simulation-based Processor Verification}
\label{cosim_report}
In co-simulation–based verification, both the HDL-written processor and a software reference model execute the same instructions, and their architectural states are compared to ensure correctness \cite{logic_fuzzer}.
The two models run in parallel, and their states are checked at each instruction commit stage.
Multiple open-source processor cores have adopted this verification framework \cite{ibex_link, cva6_link, rocket_link}.
Accordingly, we build \tool on top of this co-simulation–based framework to enable precise fault localization by leveraging fine-grained instruction-level information.
After co-simulation, an architectural state mismatch of a signal $sig$ is detected at instruction $I$ and time $t$, with the expected behavior denoted as $E$ in natural language.
The final test report for this failing instruction is represented as $(I, sig, t, E)$. 
An example is shown in Figure~\ref{fig:case}.


\vspace{-10pt}
\subsection{Fault Localization for Hardware Designs}
\label{bg:fl_hdl}

Fault localization (FL), which automatically identifies suspicious statements in source code, has been extensively studied in software systems \cite{fl_survey_tse}.
However, due to the fundamental differences between HDLs (e.g., Verilog) and software languages (e.g., Java), software-oriented FL techniques cannot be directly applied \cite{spectrum_wu}. Consequently, HDL-specific approaches have been developed.

Static slicing–based methods \cite{ahmad2022cirfix} identify suspicious signals and then perform dataflow analysis to include all statements related to these signals.
However, they often produce excessively large suspicious sets, making manual inspection difficult.
Signal transition–based methods \cite{yang2023strider} use clock timing to exclude uncovered statements, reducing the set size; however, they analyze only the cycle of the failure, limiting effectiveness in multi-cycle instruction executions.
Spectrum-based methods \cite{spectrum_wu} collect HDL coverage information from both passing and failing test cases and compute a suspiciousness score for each statement.
However, this approach heavily relies on the number of failing tests to distinguish coverage between passing and failing tests.

Recently, large language model (LLM)-based methods \cite{lik} have been proposed to predict buggy lines in Verilog modules.
These static approaches ignore dynamic processor states, which may result in limited accuracy when applied to large-scale processor designs.
Furthermore, due to the limited input context of LLMs, providing the entire project code at once is infeasible, making their application in real-world processor projects challenging.
\vspace{-5pt}
\section{Methodology}

\subsection{Overview}

\begin{figure}[htbp]
    \centering
    \includegraphics[width=\linewidth, keepaspectratio]{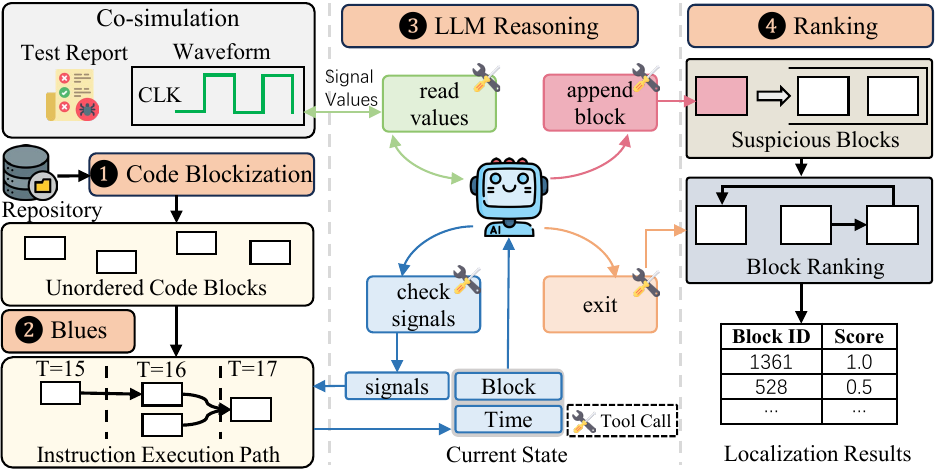}
    \caption{Overview of \tool.}
    \label{fig:workflow}
    \vspace{-10pt}
\end{figure}

\tool is a block-level fault-localization approach for processor designs that aims to locate buggy code blocks.  
As shown in Figure~\ref{fig:workflow}, after co-simulation, a test report for the failing instruction is generated.  
\ding{182} \tool first partitions the processor codebase into a set of disjoint code blocks using dataflow-based code blockization.  
\ding{183} Next, we apply a Block-Level Instruction-Oriented Slicing (\algo) algorithm to construct the instruction execution path, which chains blocks covered during instruction execution with time annotations.  
In this path, each node contains a code block and a timestamp indicating when the block is covered.
\ding{184} \tool performs debugging in a human-like manner via tool calls: at each state $(b, t)$, the LLM analyzes the current code block context to identify suspicious signals, then queries the code blocks driving these signals in the instruction execution graph.
The LLM can also leverage the time annotation to read local signal values from the waveform, aiding reasoning.  
If a block is considered suspicious, it is appended to a suspicious block queue.
\ding{185} This loop runs for multiple rounds until the LLM requests to exit the analysis. \tool then ranks all suspicious blocks by score, producing the final localization results.

\vspace{-5pt}
\subsection{Dataflow-based Code Blockization}
\label{code_blocklization}


\begin{definition}[Code Block]\label{def_code_block}
Let $L = \{ l_1, l_2, \ldots, l_m \}$ denote the set of all code lines in the HDL source code.  
A \emph{code block} $b$ is defined as a subset of lines $b \subseteq L$ such that for any two distinct blocks $b_i, b_j$, we have $b_i \cap b_j = \varnothing$.  
The set of all code blocks is denoted as $B = \{ b_1, b_2, \ldots, b_n \}$.
\end{definition}



\vspace{-5pt}
Based on Definition~\ref{def_code_block}, we propose a \textit{dataflow-based blockization} method that clusters together statements with dataflow relationships to preserve more complete local semantics.
According to the types of statements in the synthesizable subset \cite{sutherland2013synthesizing} of SystemVerilog, we categorize code blocks into four types: \textit{ModInputBlock}, \textit{ModOutputBlock}, \textit{AssignBlock}, and \textit{AlwaysBlock}.
For each block, we define an input signal set $V_i$ and an output signal set $V_o$ as follows:

\begin{itemize}[left=0pt]
    \item \textit{ModInputBlock}: $V_o$ contains only this input port signal, while $V_i$ consists of the signals connected to this input port in a module instantiation.
    \item \textit{ModOutputBlock}: $V_o$ consists of the signals connected to this output port in a module instantiation, while $V_i$ contains only this output port signal.
    \item \textit{AlwaysBlock}: $V_o$ includes all signals appearing on the left-hand side of assignments within this \texttt{always} block, and $V_i$ includes all signals appearing on the right-hand side of assignments and in conditions within the block.
    \item \textit{AssignBlock}: We first create \textit{AssignBlock} for each \texttt{assign} statement. $V_i$ is all right-hand signals in this \texttt{assign} statement and $V_o$ is all left-hand signals.
    If $V_o$ of one \textit{AssignBlock} block shares a dataflow dependency with $V_i$ of another \textit{AssignBlock} block, we merge them by combining their respective $V_i$ and $V_o$.
\end{itemize}
\vspace{-10pt}

\begin{algorithm}[htbp]
\caption{Block-Level Instruction-Oriented Slicing}
\label{alg:bios}
\KwIn{Code block set $\mathcal{B}$, initial signal $sig$, time $t$}
\KwOut{Instruction execution path $G$}
\SetKw{Continue}{continue}

Initialize a queue $S$ with $(sig, t)$\;

\While{$S$ is not empty}{
    $(s, t_{cur}) \gets \text{Pop}(S)$\;

    \If{$t_{cur} < 0$}{
        \Continue
    }
    $b \gets \text{FindDrivenBlock}(s_)$\;
    $(driven\_signals, t') \gets \text{IntraBlockAnalysis}(s, b, t_{cur})$\;

    \ForEach{$s_i \in driven\_signals$ \label{line:inter_1} }{
        $b' \gets \text{FindDrivenBlock}(s_i)$ \label{line:inter_db} \;
        Add node $(b', t')$ and edge $(b', t') \to (b, t_{cur})$ to $G$\;
        Push $(s_i, t')$ to $S$ \label{line:inter_2} \;
    }
}
\Return $G$\;

\SetKwFunction{FIntraBlockAnalysis}{IntraBlockAnalysis}
\SetKwProg{Fn}{Function}{:}{}
\Fn{\FIntraBlockAnalysis{$s, b, t$}}{
    $I_s \gets \text{DataflowAnalysis}(s, b, t)$ \label{line:df} \tcp*{Get input signals driving $s$}

    \If{$b$ is \texttt{COMB} \label{line:comb_1} }{
        \Return $(I_s, t)$ \label{line:comb_2} \;
    }
    \Else{
        \If{assignment to $s$ is covered at $t-1$ \label{line:seq_cov_1} }{
            \Return $(I_s, t - 1)$ \label{line:seq_cov_2} \;
        }
        \Else{
            \Return $(\{s\}, t - 1)$ \label{line:seq_uncov} \;
        }
    }
}
\end{algorithm}

\vspace{-20pt}
\subsection{Block-Level Instruction-Oriented Slicing}
\label{blues}

Even after clustering multiple lines into individual blocks, the number of resulting blocks remains large, making it impractical for LLMs to analyze them one by one.
To exclude blocks that are not covered during failing instruction execution, we propose Block-Level Instruction-Oriented Slicing (\algo) algorithm. 
\algo constructs the instruction execution path as a graph, where each node represents a code block annotated with a timestamp\footnote{Note that all time values in this paper are measured as the number of posedge clock cycles since the start of simulation.}.
\algo consists of two complementary components: \textit{Intra-Block Analysis}, which analyzes the dataflow of suspicious signals within blocks, and \textit{Inter-Block Analysis}, which excludes code blocks that are not covered during instruction execution.


Algorithm~\ref{alg:bios} summarizes the basic workflow of \algo. 
Starting from an initial suspicious signal $sig$ and its timestamp $t$ obtained from the test report, \algo iteratively traces backward through the blocks that drive this signal. 
At each step, \textit{Intra-Block Analysis} identifies the driven signals (i.e., the signals used to compute the suspicious signal) from the input set $V_i$ of the current block. 
\textit{Inter-Block Analysis} then finds the driven blocks, which are the blocks whose output sets $V_o$ contains these driven signals.
This process annotates each block with its corresponding timestamp, forming a node $(b, t)$ and adding dependency edges between nodes in the instruction execution path $G$. 
The recursion continues until no further expansion of $G$ is possible.
Each node $(b, t)$ in $G$ indicates that block $b$ is covered at time $t$.
Note that the same block may appear in multiple nodes with different timestamps, reflecting repeated coverage during this instruction execution.
\vspace{-5pt}

\vspace{-10pt}
\subsubsection{Intra-Block Analysis}


\begin{figure}[htbp]
    \centering
    \includegraphics[width=\linewidth, keepaspectratio]{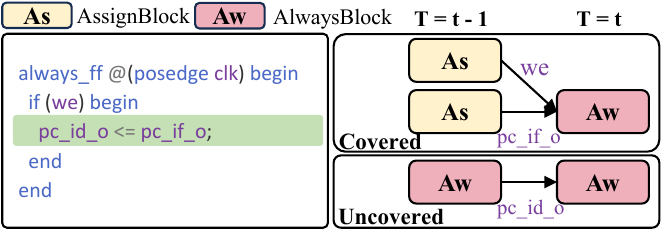}
    \caption{An example of \algo.}
    \label{fig:example}
    \vspace{-10pt}
\end{figure}

For an output signal $s \in V_o$ of block $b$ at time $t$, \textit{Intra-Block Analysis} identifies the driven signals within $V_i$ that drive $s$ by performing dataflow dependency analysis (line \ref{line:df}).
Each identified driven signal is annotated with a new timestamp $t'$, indicating that the value of $s$ at time $t$ is propagated from the values of the driven signals at time $t'$.
The timestamp $t'$ is inferred based on the type of block. 
For combinational blocks (e.g., \textit{ModInputBlock}, \textit{ModOutputBlock}, \textit{AssignBlock}, or an \textit{AlwaysBlock} without a clock), assignments are always covered, so $s$ at time $t$ is driven by the values of the driven signals at the same time $t$ (line \ref{line:comb_1}-\ref{line:comb_2}).
For sequential blocks (e.g., an \textit{AlwaysBlock} with a clock), if the assignment to $s$ is covered at time $t-1$, indicating that the value of $s$ at time $t$ is computed by the driven signals at time $t-1$, then these driven signals at time $t-1$ are collected for backward analysis (line \ref{line:seq_cov_1}-\ref{line:seq_cov_2}).
If the assignment is not covered at $t-1$, the register retains its previous value, so $s$ at time $t$ holds its own value at time $t-1$ (line \ref{line:seq_uncov}).
These driven signals and associated timestamps are provided to the LLM when it inspects signal $s$ in block $b$, ensuring that it focuses only on relevant signals and reducing interference from unrelated ones.

\vspace{-5pt}
\subsubsection{Inter-Block Analysis}
After identifying the driven signals of $s$ in block $b$ at time $t$, \textit{Inter-Block Analysis} determines which blocks actually compute the values of these signals (i.e., driven blocks) and inserts the corresponding predecessor nodes for $(b, t)$ in the instruction execution path $G$ (line \ref{line:inter_1}-\ref{line:inter_2}). Specifically, for each driven signal annotated with timestamp $t'$ in block $b$ through \textit{Intra-Block Analysis}, the block whose output signal set contains this driven signal is identified as a driven block $b'$, forming the corresponding node $(b', t')$. Then, this node is inserted into the path $G$, adding an edge from $(b', t')$ to $(b, t)$.
When the LLM identifies suspicious signals (i.e., a subset of driven signals) in block $b$ and wants to check code blocks that compute these signals, it can request the instruction execution path $G$ to get block $b'$.

As shown in Figure \ref{fig:example}, if the green line is covered at time $t-1$, driven signals of \texttt{pc\_id\_o} include both \texttt{pc\_if\_o} and \texttt{we} at $t-1$. Then, after \textit{Inter-Block Analysis}, two \textit{AssignBlock}s that drive these signals are connected to this \textit{AlwaysBlock}.
If this line is not covered, the register \texttt{pc\_id\_o} retains its value, meaning it is driven by itself from time $t-1$ through the same \textit{AlwaysBlock}.

\vspace{-10pt}
\subsection{LLM Reasoning and Ranking}
\label{sec:reasoning}

\vspace{-10pt}
\begin{figure}[htbp]
    \centering
    \includegraphics[width=\linewidth, keepaspectratio]{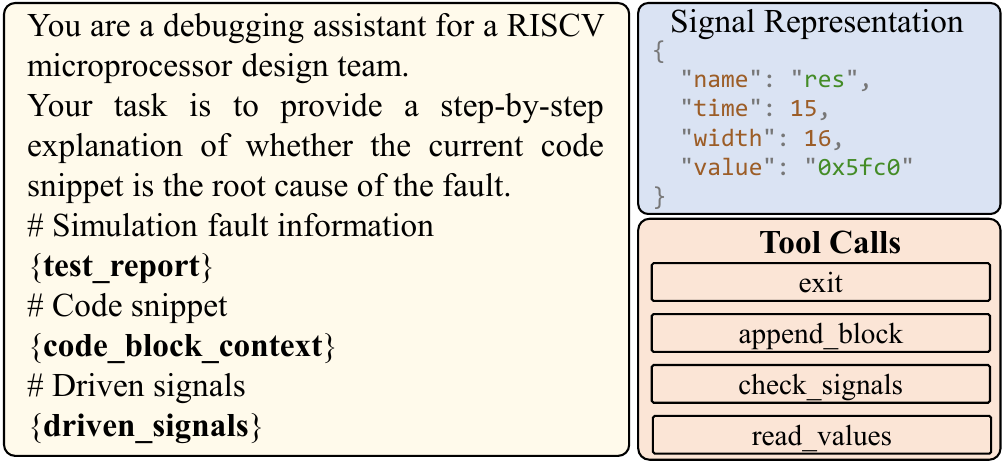}
    \caption{The prompt template of \tool.}
    \label{fig:prompt}
\end{figure}

\vspace{-13pt}
We prompt LLM reasoning and make decisions via tool calls at each state $(b, t)$. 
As shown in Figure~\ref{fig:prompt}, the LLM evaluates whether current code block context is the root cause of the bug.
If this block is deemed suspicious, it calls the \textit{append\_block} tool to append $b$ to suspicious block queue.
Driven signals obtained by \textit{Intra-Block Analysis} of \algo are provided in the prompt as a JSON array, with each signal represented in the format shown in Figure \ref{fig:prompt}.
If LLM requires further inspection, it queries a set of suspicious signal names via the \textit{check\_signals} tool.
All signals selected by the LLM must be chosen from the provided list to ensure the LLM focuses only on blocks within the execution path.
\algo maps these signals to their driven blocks via the instruction execution path $G$, generating new states for the LLM to inspect.
With timestamps annotated for each block, the LLM can retrieve critical signal values from the waveform using the \textit{read\_values} tool, aiding in instruction behavior comprehension.
This loop executes for multiple rounds until the LLM confirms that the root cause is in the suspicious queue and calls the \textit{exit} tool. It then assigns each suspicious block a confidence score between 0 and 1, producing a ranked list of suspicious blocks.
\vspace{-15pt}




\section{Experiment}

\subsection{Experiment Setup}

\subsubsection{Benchmark}

We adopt a production-quality, open-source RISC-V core Ibex \cite{ibex_link} to evaluate \tool. It was selected for its wide adoption in the open-source hardware community (1.7K GitHub stars) and its implementation in approximately 19 KLoC of SystemVerilog, providing a realistic scale for evaluating \tool's performance on real-world processor projects.
Following prior work \cite{lik, sudakrishnan2008understandingverilog}, we inject bugs into Ibex according to the same mutation rules. The \textit{CoreMark} \cite{coremark_link} test program
 is executed in a co-simulation framework to trigger the injected bugs. If verification fails, we record the buggy line and save the modified code. \textbf{In total, 119 bugs are injected to construct our benchmark.}

\vspace{-5pt}
\subsubsection{Baselines}

We compare \tool with state-of-the-art FL techniques for HDLs, including Cirfix~\cite{ahmad2022cirfix} (static slicing-based), Strider~\cite{yang2023strider} (transition-based), Tarsel~\cite{spectrum_wu} (spectrum-based), and LiK~\cite{lik} (LLM-based).
LiK requires both the module code and its specification; therefore, we use \textit{GPT-4} to summarize Ibex modules into concise specifications following the LiK dataset construction methodology.
\textbf{As LiK localizes bugs only within a single module, we manually provide the buggy module file for project-level evaluations.}
While this gives LiK additional prior knowledge over \tool, it ensures an available comparison.

We also implement a LLM-based software-oriented FL approach, \naive \cite{agentless}, adapted for processors. Following its repository-oriented workflow, \naive first uses an LLM to identify potentially buggy module files from the test report and project structure, then provides each suspicious file's full source code to the LLM to predict bug locations.


\vspace{-5pt}
\subsubsection{Implementation Details}

The \tool prototype utilizes the \texttt{sv-parser}\cite{sv_parser} library to parse the abstract syntax tree (AST) of SystemVerilog files.
We employ Verilator\cite{verilator} to perform simulations and generate waveforms for collecting signal values and code coverage information. 
\tool is implemented in approximately 12 KLoC of Rust, which includes the \algo algorithm and LLM-based reasoning components.
\tool uses \textit{GPT-4o} model as default.

\vspace{-5pt}
\subsubsection{Evaluation Metrics}

Following prior studies in the FL field~\cite{spectrum_wu, autofl}, we evaluate \tool using the Top-$N$ metric, which counts the number of bugs for which at least one buggy block appears within the top $N$ positions of the ranked list. Slicing-based approaches output only sets of statements without suspiciousness scores. Following~\cite{spectrum_wu}, we assign each statement an expected rank of $|S|/2$, where $|S|$ is the set size.
LiK outputs a single buggy line, so its Top-$N$ results remain constant across $N$. For fair comparison, all baseline results are aligned to the block level.
\vspace{-5pt}

\subsection{Experiment Results}


\subsubsection{RQ1. Effectiveness of \tool}

\vspace{-10pt}
\begin{table}[htbp]
\centering
\caption{Comparison between \tool and other FL baselines on benchmark.}
\label{tab:rq1}
\begin{tabular}{cc|ccc}
    \toprule
    \textbf{Approach} & \textbf{Type} & \textbf{Top-1} & \textbf{Top-5} & \textbf{Top-10} \\ 
    \midrule
    Cirfix            & Project & 0  & 0  & 0 \\
    Strider           & Project & 0  & 0  & 0 \\
    Tarsel-Jaccard    & Project & 0  & 0  & 0 \\
    Tarsel-Ochiai     & Project & 0  & 0  & 2 \\
    Tarsel-Tarantula  & Project & 3  & 4  & 7 \\
    LiK & Module  & 7  & 7  & 7 \\
    \scshape{Agentless}            & Project & 13  & 16  & 16 \\
    \midrule
    \textbf{\tool}                  & Project & \textbf{24} & \textbf{28} & \textbf{28}\\
    \ding{182} w/o Signal Values    & Project & 13 & 24 & 26 \\
    \ding{183} w/o Instruction Path & Project & 5 & 6 & 6 \\
    \bottomrule
\end{tabular}
\vspace{-10pt}
\end{table}

Table~\ref{tab:rq1} summarizes the overall effectiveness of \tool compared with existing FL baselines. \tool localizes 24, 28, and 28 bugs within the Top-1, Top-5, and Top-10 rankings, respectively, outperforming the best existing baseline, LiK, by up to 242.9\% in Top-1 accuracy.

Slicing-based approaches generate overly large suspicious sets, hindering accurate Top-10 localization.
Among spectrum-based FL metrics, Tarantula performs best, but its overall accuracy remains limited.
These methods rely on range-based coverage collection over simulation intervals, causing coverage data for a failing instruction to be polluted by other in-flight instructions.

\begin{figure}[htbp]
    \centering
    \includegraphics[width=\linewidth, keepaspectratio]{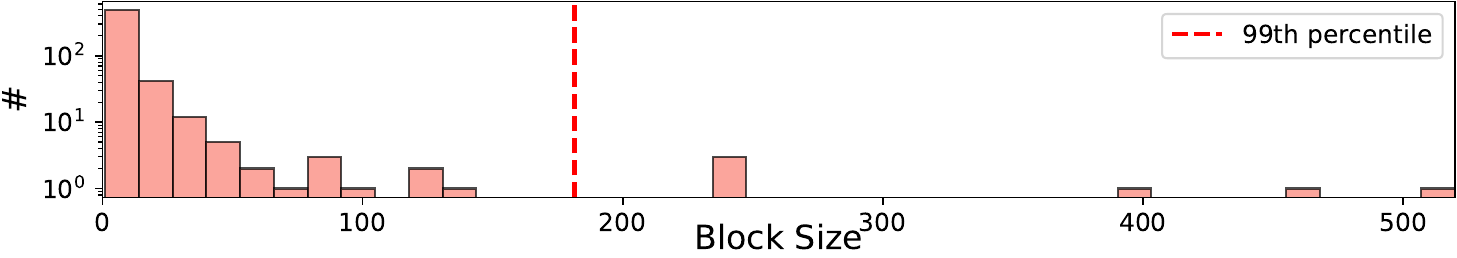}
    \caption{Histogram of Block Sizes in Ibex (Log Y Scale).}
    \label{fig:block_sizes}
    \vspace{-10pt}
\end{figure}

For the LLM-based, hardware-specific FL approach LiK, only 7 bugs are localized, as the long module code in Ibex hinders precise locating buggy lines.
\naive provides entire module file to \textit{GPT-4o} and also shows limited results compared to \tool due to excessively long code context.  
In contrast, \tool employs dataflow-based code blockization to retain compact yet relevant context.
As shown in Figure~\ref{fig:block_sizes}, 99\% of code blocks parsed by our approach for Ibex contain no more than 200 lines, except for a few special modules such as the instruction decoder.  
Moreover, \algo guides LLMs to reason only over blocks covered during instruction execution and leverage signal values to comprehensively understand instruction behavior, resulting in the best performance among baselines.
\vspace{-10pt}

\vspace{-15pt}

\subsubsection{RQ2. Effectiveness of \algo}

\begin{figure}[htbp]
    \centering
    \includegraphics[width=\linewidth, keepaspectratio]{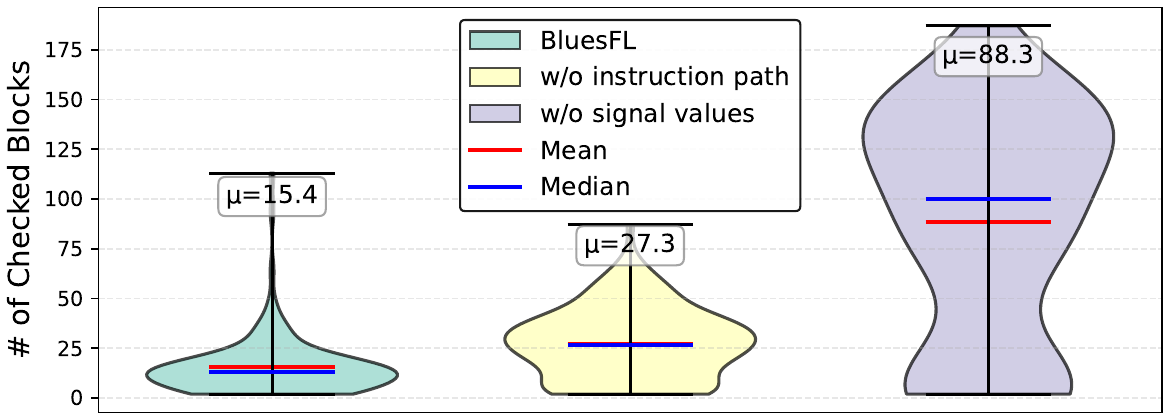}
    \caption{Distribution of the number of checked blocks across different settings.}
    \label{fig:trace_size_dis}
    \vspace{-10pt}
\end{figure}

\algo guides LLMs to exclude irrelevant code blocks by following the instruction execution path and helps them understand processor behavior through signal values.
We conduct an ablation experiment to evaluate the contribution of signal values and execution paths in localizing bugs.
In setting \ding{182}, we remove \textit{read\_values} from tools given to \tool, preventing the LLM from accessing waveform.  
In setting \ding{183}, we remove the instruction execution path obtained by \algo, allowing the LLM to freely access any code blocks.

As shown in Table \ref{tab:rq1}, both \ding{182} and \ding{183} show a performance drop compared with \tool, indicating that both signal values and instruction execution paths are necessary for effective processor debugging.  
Specifically, without the instruction execution path, LLMs often access blocks not covered during instruction execution, focusing on irrelevant blocks and producing false positives.  
Interestingly, we observe that without signal values, LLMs tend to access more code blocks. As shown in Fig.~\ref{fig:trace_size_dis}, the average number of blocks checked by \tool increases to 88.3, compared to 15.4 for \tool, resulting in a 609\% increase in cost per bug (\$0.257 $\rightarrow$ \$1.821).
This indicates that, without signal values, LLMs require more code context to fully reason about the bug.

\subsubsection{RQ3: Performance on Other LLMs}

We evaluate \tool across three LLM backends: \textit{GPT-4o-mini}, \textit{GPT-4o}, and \textit{Claude-3.5-Sonnet}. 
\textit{GPT-4o-mini} provides a low-cost option, whereas \textit{GPT-4o} and \textit{Claude-3.5-Sonnet} offer stronger code capabilities.
As shown in Table~\ref{tab:rq3}, \tool performs competitively even with the inexpensive \textit{GPT-4o-mini}, and achieves higher performance with \textit{Claude-3.5-Sonnet} at increased cost.

\vspace{-8pt}
\begin{table}[htbp]
    \centering
    \caption{Performance of \tool on Different LLMs.}
    \label{tab:rq3}
    \begin{tabular}{l|cccccc}
    \toprule
        \textbf{Base Model} & \textbf{Top-1} & \textbf{Top-5} & \textbf{Top-10} & \textbf{Avg Cost} \\ 
        \midrule
        GPT-4o-mini         & 17 & 25 & 26 & \$0.015\\
        GPT-4o              & 24 & 28 & 28 & \$0.257\\
        Claude-3.5-Sonnet   & 30 & 33 & 33 & \$0.263\\
    \bottomrule
    \end{tabular}
\end{table}
\vspace{-20pt}

\section{Discussion}

\subsection{How \tool works}


To better understand \tool's debugging behavior, we visualize how it localizes a bug in Ibex.
We injected a bug in \texttt{alu} module.
Test report showed that the instruction \texttt{j} jumped to \texttt{0x000f5fc0} instead of the expected \texttt{0x0010a140}.
Without prior knowledge of the processor, linking this symptom to the \texttt{alu} module is not obvious.

\begin{figure}[tbp]
    \centering
    \includegraphics[width=\linewidth, keepaspectratio]{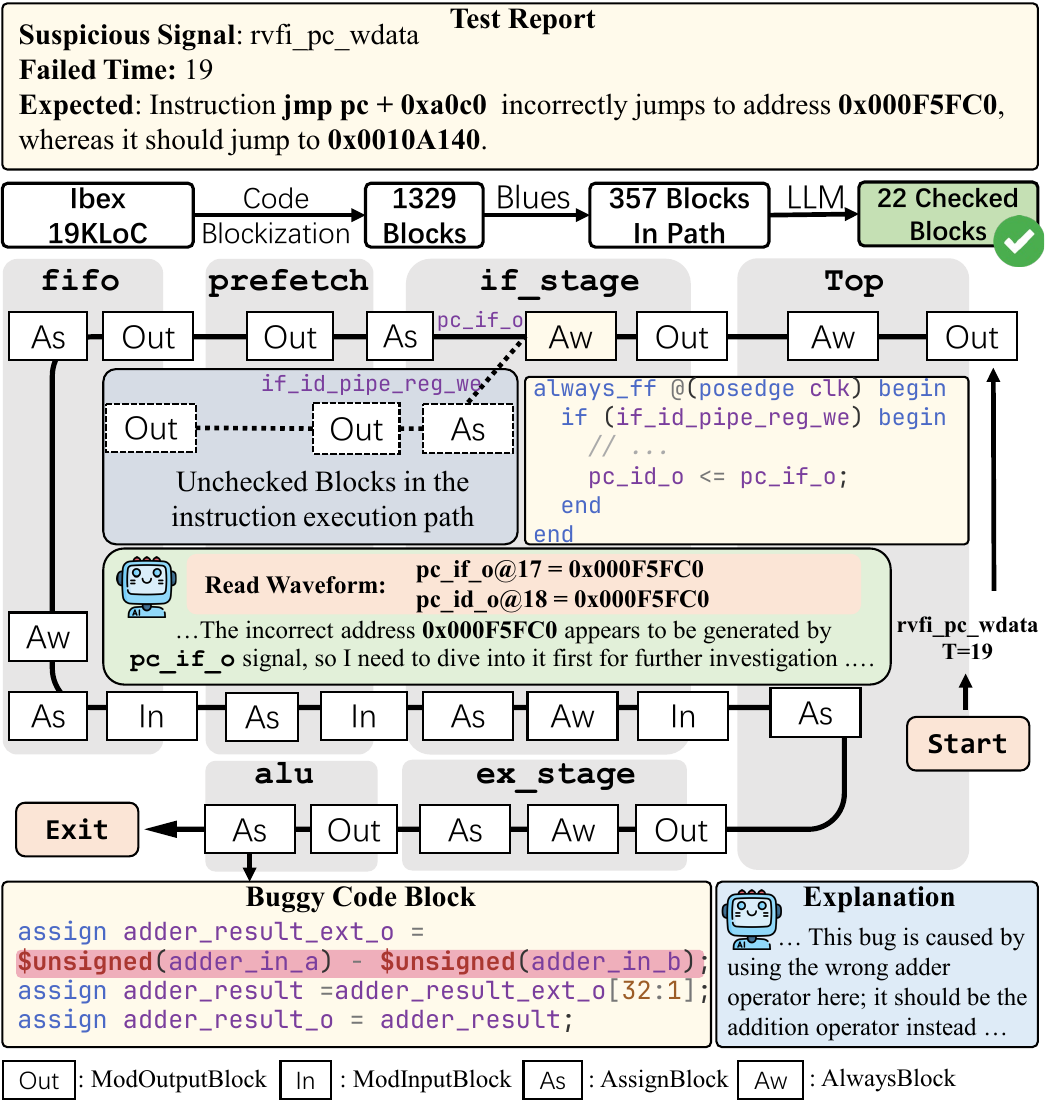}
    \caption{A case to show how \tool works.}
    \label{fig:case}
    \vspace{-10pt}
\end{figure}

As shown in Figure \ref{fig:case}, \tool first applies code blockization to partition the Ibex project into 1,329 blocks.
Then, after running \algo, the instruction execution path contains only 357 blocks.
Despite this significant reduction, the result remains challenging for humans to review.
With LLM integration, \tool inspected only 22 blocks to pinpoint the bug, as shown in Figure \ref{fig:case}.

Specifically, \tool begins its analysis from the \texttt{Top} module at time 19.
For clarity, we visualize only the LLM's behavior when inspecting an \textit{AlwaysBlock} in the \texttt{if\_stage} module.
After \textit{Intra-Block Analysis} of \algo, the LLM knows that \texttt{pc\_id\_o@18} is driven by \texttt{pc\_if\_o@17} and \texttt{if\_id\_pipe\_reg\_we@17}.
Without additional information, either signal could be the source of the bug.
However, after examining these driven signal values, the LLM notices that \texttt{pc\_if\_o@17} matches the incorrect jump address.
Consequently, it prioritizes checking \texttt{pc\_if\_o@17}, leaving the blocks that drive \texttt{if\_id\_pipe\_reg\_we@17} unchecked.
After inspecting all fetch related modules under \algo's guidance and finding no bugs, the LLM returns to the \texttt{Top} module and continues into the execution-stage modules, where it locates the bug in an \textit{AssignBlock} within the \texttt{alu} module.
By combining the test report with code context, the LLM pinpoints the buggy block and provides a natural-language explanation.
Overall, \tool traces long-distance signal propagation across space and time, behaving like a human debugger as it gathers debugging information and reasons about the root cause.


\vspace{-5pt}

\subsection{Limitation of \tool}

\begin{figure}[htbp]
    \centering
    \includegraphics[width=\linewidth, keepaspectratio]{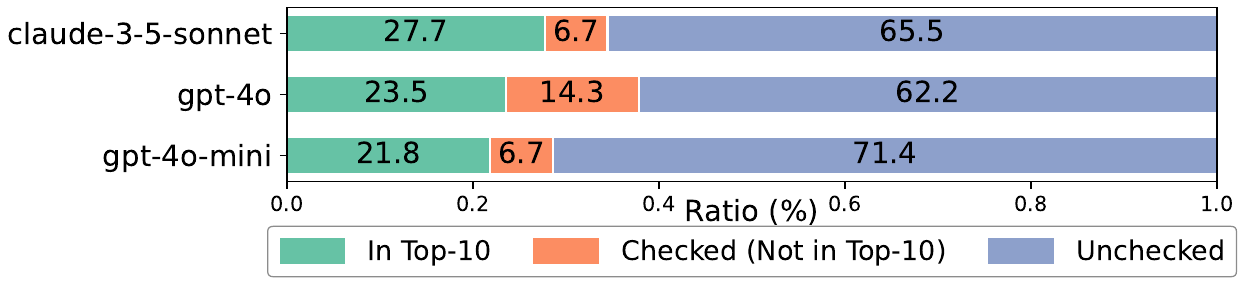}
    \caption{Ratio of Buggy Blocks Checked by \tool.}
    \label{fig:hit_ratio}
    \vspace{-16pt}
\end{figure}

Evaluation of existing FL techniques on real-world processor designs indicates that fault localization for processors is still at an early stage.
We analyzed cases in which \tool failed to localize bugs within the Top-10 across different models.
As shown in Figure \ref{fig:hit_ratio}, for approximately 70\% of these bugs, buggy blocks were never checked by LLMs because the debugging process terminated too early.
To further improve FL performance, future work could explore methods that encourage LLMs to inspect more code blocks within a project, enabling long-distance, cross-block reasoning and more comprehensive localization.
\vspace{-10pt}


\subsection{Threats to Validity}

LLMs are trained on large-scale codebases from GitHub and may have been exposed to the correct processor design code used in our evaluation, potentially causing data leakage.
To mitigate this, we inject bugs to ensure that the LLM has not encountered these test cases before.
And the poor performance of the default \textit{GPT-4o} (i.e., the baseline \naive) indicates that the model has not simply memorized the correct code, while the significant improvement of \tool over \naive demonstrates the effectiveness of our approach.
\vspace{-10pt}




\section{Conclusion}

We present a novel LLM-based fault-localization framework for processor designs.
Inspired by human debugging, \tool analyzes instruction execution paths, inspects processor states, and identifies the most suspicious code blocks.
Our dataflow-based blockization reduces code context, allowing LLMs to focus on local dataflow.
We further introduce a block-level instruction-oriented slicing algorithm (\algo) that chains instruction-covered blocks with time annotations.
\algo enables LLMs to efficiently trace instruction execution and read signal values from large waveforms to reason about instruction behavior.
Experimental results on a real-world RISC-V processor core show that \tool localizes 24 bugs at Top-1, achieving a 242.9\% improvement over the best existing baseline.
\vspace{-20pt}

\begin{acks}
This work was supported by the National Natural Science Foundation of China (No.62474196, No.62402515, and No.62504255). 
\end{acks}

\bibliographystyle{ACM-Reference-Format}
\bibliography{references}

\end{document}